\documentclass[
 reprint,
superscriptaddress,
 amsmath,amssymb,
 aps,
]{revtex4-2}

\usepackage{graphicx}
\usepackage{dcolumn}
\usepackage{bm}
\usepackage{xcolor}

\begin{document}

\preprint{APS/123-QED}

\title{The magnetic structure of polar $G$-type charge and orbital ordered Hg-quadruple manganite perovskites} 

\author{Ben R. M. Tragheim}
\affiliation{Department of Chemistry, University of Warwick, Gibbet Hill, Coventry, CV4 7AL, UK}
\affiliation{School of Chemical, Materials and Biological Engineering, University of Sheffield, Mappin Street, Sheffield, S1 3JD, UK}

\author{Fabio Orlandi}
\affiliation{ISIS Neutron and Muon Source, STFC Rutherford Appleton Laboratory, Harwell campus, Didcot, Oxfordshire, OX11 0QX, UK}

\author{En-Pei Liu}
\affiliation{Department of Physics, National Taiwan University, Taipei 10617, Taiwan}
\affiliation{Center for Condensed Matter Sciences, National Taiwan University, Taipei 10617, Taiwan}

\author{Wei-Tin Chen}
\email{weitinchen@ntu.edu.tw}
\affiliation{Center for Condensed Matter Sciences, National Taiwan University, Taipei 10617, Taiwan}
\affiliation{Center of Atomic Initiative for New Materials, National Taiwan University, Taipei 10617, Taiwan}
\affiliation{Taiwan Consortium of Emergent Crystalline Materials, National Science and Technology Council, Taipei 10622, Taiwan}

\author{Mark S. Senn}
\email{m.senn@warwick.ac.uk}
\affiliation{Department of Chemistry, University of Warwick, Gibbet Hill, Coventry, CV4 7AL, UK}

\date{\today}

\begin{abstract}

The magnetic structure of the novel Hg$_{0.7}$Na$_{0.3}$Mn$_3$Mn$_4$O$_{12}$, a quadruple manganite perovskite that exhibits a unique $G$-type charge and orbital ordered state distinct to other $A^{2+}$Mn$_3$Mn$_4$O$_{12}$ equivalents ($A$ $=$ Ca, Sr, Cd, Pb), has been solved using powder neutron diffraction and symmetry-motivated analysis. A $G$-type-like antiferromagnetic (AFM) ordering of Mn on the $A'$ sites and a `up--up--down--down' AFM moment configuration of Mn spins on the $B$ sites is found to occur. The mechanism for the onset and stabilization of $B$ site `up--up--down--down' AFM order is explored in terms of coupling between structural and magnetic distortions. The results presented here provide evidence of the exotic charge, orbital, electronic and magnetic orderings that quadruple manganite perovskites demonstrate, and further highlighting the distinct chemistry that Hg$^{2+}$ plays in stabilizing novel states compared to other divalent $A$-site cation equivalents. 
 
\end{abstract}

\maketitle

\section{Introduction}

The emergence of intricate couplings between structural, electronic and magnetic degrees of freedom are ubiquitously observed in the $A$MnO$_3$ manganite perovskites ($A$ $=$ $A$-site cation), notably resulting in the formation of functionally important properties such as colossal magnetoresistance (CMR)\cite{jin1994thousandfold, uehara1999percolative} and magnetoelectric multiferroism.\cite{kimura2003magnetic} Quadruple manganite perovskites $A$Mn$_3$Mn$_4$O$_{12}$, of the structural form $AA'_3B_4$O$_{12}$, are unique derivatives of the $AB$O$_3$ perovskites that have had recent significant research interest due to their ability to stabilize additional exotic charge, orbital, magnetic and electronic orderings compared to their $AB$O$_3$ counterparts.\cite{chen2018improper, chen2021striping, Tragheim134, perks2012magneto, khalyavin2020emergent, johnson2017magneto, lu2024giant} These different orders and systems then provide alternative avenues through which the coupling between different degrees of freedom may occur by in perovskite-related systems, and hence the ability to design and optimize novel functional materials. 

One particular subset of quadruple manganite perovskites include those with a divalent $A$ site cation ($A$ $=$ Ca, Sr, Cd, Pb, Hg)\cite{bochu1974high, locherer2012synthesis, chen2018improper}. In these, Mn in the square planar $A'$ site is generally robust to adopt a $+$3 valence state, and Mn in the octahedral $B$ sites in general adopt a 3:1 $+$3:$+$4 valence state ratio, giving an average Mn $B$ site oxidation state, $<$Mn$_B$$>$, of $+$3.25. At high temperatures each of these adopt the $Im\bar{3}$ space group, fully charge and orbital disordered structure. Upon cooling each undergoes a phase transition to an $R\bar{3}$ space group, charge ordered but orbital disordered state referred to as `disordered $C$-type'\cite{Tragheim134}, structure. A successive low temperature structural phase transition occurs in each. For $A$ $=$ Ca and Sr, an incommensurate $R\bar{3}$(00$\gamma$)0 orbital density wave (ODW) ground state occurs, and for $A$ $=$ Cd the ODW is commensurate and a $P\bar{3}$ space group forms instead\cite{johnson2017magneto}. $A$ $=$ Pb is more complex in comparison as an initial incommensurate $R\bar{3}$(00$\gamma$)0 ODW state forms, followed by a quasicommensurate structural modulation transition, and then a re-entrant transition occurs back to the incommensurate $R\bar{3}$(00$\gamma$)0 ODW state upon cooling\cite{johnson2021competing}. However, for $A$ $=$ Hg the incommensurate $R\bar{3}$(00$\gamma$)0 state does not form, but instead the polar $P$2 (with $Pnn$2 pseudosymmetry) space group structure forms, demonstrating a novel $G$-type charge and orbital ordered state not observed before in any manganite perovskites known to date\cite{chen2018improper, Tragheim134}. The effect of electronic hole hoping $via$ substitution of the divalent $A$ site cation for Na$^{+}$, notably observed in the solid solution Ca$_{1-x}$Na$_x$Mn$_3$Mn$_4$O$_{12}$, also influences the ground state charge and orbital order these quadruple perovskites adopt\cite{johnson2018evolution, chen2021striping}. This is illustrated by the formation of the novel orbital order:charge disorder (OO:CD)-type state at optimal doping with $x$ $\approx$ 0.5, corresponding to $<$Mn$_B$$>$ $\approx$ $+$3.375\cite{chen2021striping}. Here, a monoclinic $C$2/$m$ space group structure is adopted, which again is distinctly different to the incommensurate $R\bar{3}$(00$\gamma$)0 states that the majority of divalent quadruple manganite perovskites adopt. 

The stark differences between the space group symmetries and charge/orbital orderings of these systems may also be demonstrated in the differences of the magnetic structures they adopt. For quadruple manganite perovskites with $A$ $=$ Ca, Sr, Cd and Pb the magnetic structures have been well studied\cite{johnson2016modulated, johnson2017magneto}. For $A$ $=$ Ca, Sr and Pb an initial antiferromagnetic (AFM) transition occurs which produces an incommensurate spin density wave where orbital and magnetic modulations lock together through a commensurate magneto-orbital coupling. At lower temperatures a second AFM transition occurs which delocks the orbital and magnetic modulations from each other, resulting in an incommensurate magneto-orbital coupling. In this state the magnetic moments form a chiral and polar constant moment helix with a modulated spin helicity. For $A$ $=$ Cd this same incommensurate magneto-orbital coupling ground state occurs, but the higher temperature AFM transition instead forms a commensurate magnetic structure\cite{johnson2017magneto}. The magnetic structures of hole doped variants of these quadruple perovskites are also subject to change as well. For example, hole doped Ca$_{1-x}$Na$_x$Mn$_3$Mn$_4$O$_{12}$ for $x$ $\approx$ 0.5 are shown to exhibit a unique incommensurate pseudo $CE$-type AFM structure\cite{johnson2018evolution}, closely related to the commensurate pseudo $CE$-type AFM structures exhibited by Pr$_{1-x}$Ca$_x$MnO$_3$ in the same nominal $<$Mn$_B$$>$ range\cite{jirak1985neutron,yoshizawa1995neutron}.

With these points in mind, we aim to solve the magnetic structure of Hg-based $A$ site quadruple manganite perovskites that adopt the novel $G$-type charge and orbital ordered ground state. For the present study we investigate the composition Hg$_{0.7}$Na$_{0.3}$Mn$_3$Mn$_4$O$_{12}$ since in our previous work on our samples of Hg$_{1-x}$Na$_{x}$Mn$_3$Mn$_4$O$_{12}$ it is this composition that contains the lowest amount of a secondary $R\bar{3}$ phase, containing minimal phase coexistence, which otherwise precludes an accurate determination of the magnetic structure attributed to the $G$-type charge and orbital ordered state. We find that Hg$_{0.7}$Na$_{0.3}$Mn$_3$Mn$_4$O$_{12}$ exhibits two AFM transitions at $T_\text{N}$ $\approx$ 80\,K and 60\,K due to ordering of the $B$ and $A'$ sites, respectively. The resulting magnetic unit cell is found to be double that of the $P$2 crystal structure, with a doubling of the Jahn--Teller (JT) long axis direction in which $G$-type orbital order, tetragonal strain with respect to the $Im\bar{3}$ $AA'_3B_4$O$_{12}$ aristotype and ferroelectric polar distortions occur along. By testing different symmetry-allowed magnetic structure models we find that only a model in $P2_{1}^{'}$ symmetry (in Belov-Neronova-Smirnova (BNS) notation) can satisfactorily provide a physically-sensible model. This magnetic structure is found to demonstrate a $G$-type-like AFM order of $A'$ site moments, and an `up--up--down--down' AFM order configuration of $B$ site moments aligned perpendicular to $A'$ site moments. A slight canting of $B$ site moments is also found to occur along a concomitant direction with that of the predisposed $B$O$_6$ octahedral rotations. 

\section{Experiment and data analysis}

Synthesis details of Hg$_{0.7}$Na$_{0.3}$Mn$_3$Mn$_4$O$_{12}$ can be found in our previous work\cite{Tragheim134}. Magnetic susceptibility data were collected over a temperature range 5\,K $\leq$ $T$ $\leq$ 300\,K under field cooled (FC), at 1000\,Oe, and zero field cooled (ZFC) conditions using a Quantum Design MPMS3 magnetometer. Time-of-flight powder neutron diffraction (PND) data were collected on a combined $\sim$ 40\,mg sample on the  WISH\cite{chapon2011wish} beamline at the ISIS Neutron and Muon Source (UK). Data were collected on warming at $T$ $=$ 1.5\,K, 10\,K $\leq$ $T$ $\leq$ 100\,K in 10\,K increments, 250\,K. Data at $T$ $=$ 1.5\,K, 40\,K and 100\,K were collected for 3.75\,h (140\,$\mu$A) and all other data were collected for 30 minutes (20\,$\mu$A). 

The generation of all symmetry-allowed magnetic structure models, and all symmetry analysis, were performed using the software ISODISTORT\cite{Campbell1} and INVARIANTS\cite{Hatch1} contained within the ISOTROPY suite\cite{Isotropy}. For symmetry analysis, structural and magnetic distortions of each model transform as irreducible representations (irreps) with respect to an $Im\bar{3}$ $AA'_3B_4$O$_{12}$ quadruple perovskite aristotype with the $A$ site placed at the origin. Structural models generated throughout the present work use the standard setting of monoclinic space groups (unique \textbf{\textit{b}} axis, cell choice 1) compared to the non-standard setting (unique \textbf{\textit{c}} axis, cell choice 2) used for the crystal structure previously given\cite{Tragheim134}. Hence, with respect to the $P$2 structure, in the present work the direction of the long Mn-O JT bond in Mn$^{3+}$O$_6$ octahedra lies along \textbf{\textit{b}} rather than \textbf{\textit{c}}. This was chosen to prevent additional complications in solving a magnetic structure using non-standard settings. All Pawley and Rietveld refinements of magnetic structure models against PND data, using the software Topas Academic V7\cite{Coelho1}, refine magnetic moment amplitudes directly as imposed and guided by the symmetry analysis outlined prior. 

\section{Results and discussion}

\subsection{Magnetic susceptibility}

\begin{figure}
\includegraphics{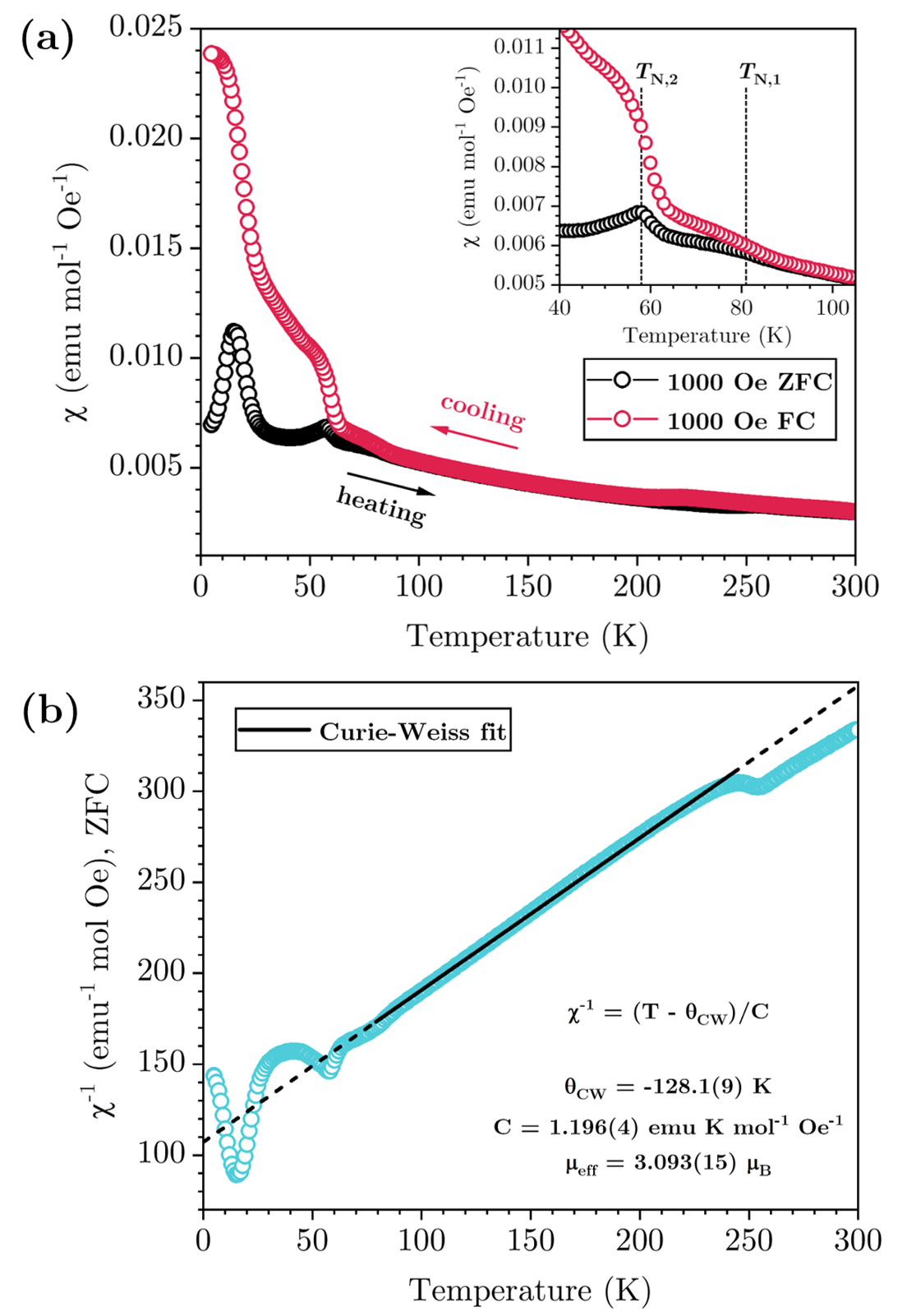}
\caption{\label{magnetic-susceptibility} (a) Magnetic susceptibility ($\chi$) measurements under ZFC and FC conditions at 1000\,Oe. The inset indicates the presence of hysteresis consistent with the $R\bar{3}$ to $P$2 structural phase transition. (b) Curie-Weiss fit to $\chi^{-1}$ ZFC data within the paramagnetic temperature range 80\,K $\leq$ $T$ $\leq$ 240\,K.} 
\end{figure}

Magnetic susceptibility data in Figure \ref{magnetic-susceptibility}(a) under ZFC and FC conditions demonstrate the presence of two inflections corresponding to AFM transitions at temperatures $T_\text{N,1}$ $\approx$ 80\,K and $T_\text{N,2}$ $\approx$ 60\,K. We attribute these transitions to a separate ordering of $B$ and $A'$ sites of Hg$_{0.7}$Na$_{0.3}$Mn$_3$Mn$_4$O$_{12}$. Modeling powder neutron diffraction in Section \ref{Powder-Neutron-Diffraction} will corroborate this. It is noted that a third inflection occurs at $T$ $\approx$ 15\,K which we attribute to $\alpha$-Mn$_2$O$_3$ impurities, present in our sample of only $\sim$ 3$\%$ weight fraction by Rietveld refinement. At higher temperatures, hysteresis is observed upon cooling and heating between 220\,K and 250\,K, indicative of, and consistent with, the first order phase transition between $R\bar{3}$ and $P$2 phases observed for all Hg$_{1-x}$Na$_{x}$Mn$_3$Mn$_4$O$_{12}$ samples studied previously\cite{Tragheim134}. An inverse Curie-Weiss equation was fit to the paramagnetic regime in the range 80\,K $\leq$ $T$ $\leq$ 240\,K, shown in  Figure \ref{magnetic-susceptibility}(b). This gives a Curie-Weiss temperature $\theta_\text{CW}$ $=$ -128.1(9)\,K indicating predominant AFM moment interactions, and a Curie constant C $=$ 1.196(4)\,emu\,K\,mol$^{-1}$\,Oe$^{-1}$ corresponding to an effective average Mn magnetic moment $\mu_\text{eff}$ $=$ 3.093(15)\,$\mu_\text{B}$. This value will encompass the range of Mn oxidation states Mn$^{2+}$, Mn$^{3+}$ and Mn$^{4+}$ as each of these are present in Hg$_{0.7}$Na$_{0.3}$Mn$_3$Mn$_4$O$_{12}$ across both the magnetically-active $A'$ and $B$ site sub-lattices.     

\subsection{Powder neutron diffraction}\label{Powder-Neutron-Diffraction}

\begin{figure}
\includegraphics{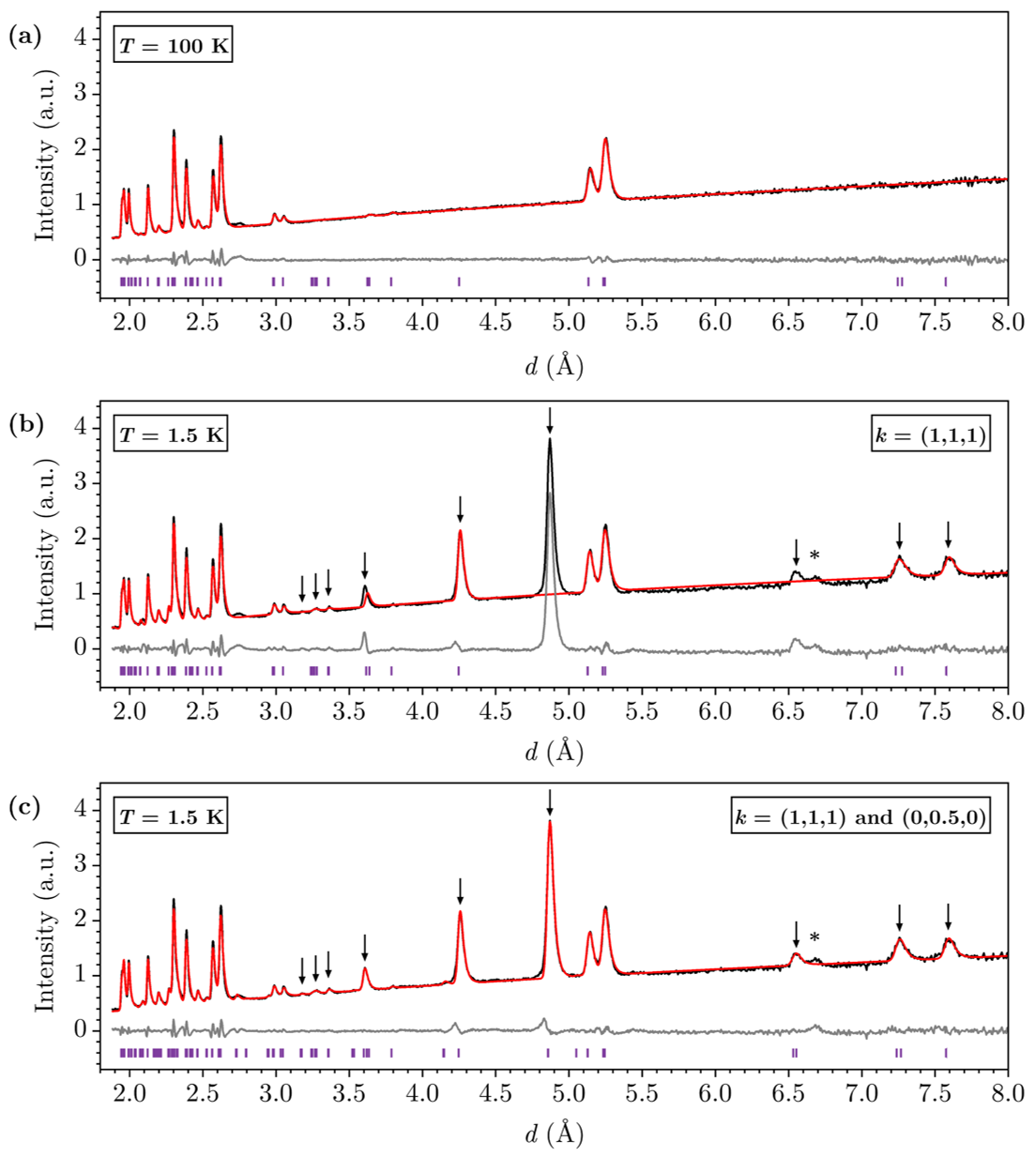}
\caption{\label{pawley-refinement} Pawley refinements of models against (a) 100\,K PND data and 1.5\,K PND data with propagation vectors (b) \textbf{\textit{k$_1$}} $=$ (1,1,1) and (c) \textbf{\textit{k$_1$}} $=$ (1,1,1) and \textbf{\textit{k$_2$}} (0,0.5,0) with respect to the $Im\bar{3}$ quadruple perovskite structure. Arrows denote magnetic peaks due to Hg$_{0.7}$Na$_{0.3}$Mn$_3$Mn$_4$O$_{12}$. The asterisk in (b) and (c) indicate the presence of a magnetic impurity, attributed to $\alpha$-Mn$_2$O$_3$. Purple ticks are peak positions of the Pawley refined structural model.} 
\end{figure}

Pawley refinements were performed against PND data collected at 1.5\,K to determine the propagation vector(s) of the magnetic ground state. These refinements demonstrate that two magnetic propagation vectors: \textbf{\textit{k$_1$}} $=$ (1,1,1) and \textbf{\textit{k$_2$}} $=$ (0,0.5,0), with respect to the $Im\bar{3}$ quadruple perovskite structure, are required to index all the magnetic peaks (Figure \ref{pawley-refinement}). These correspond to \textbf{\textit{k$_{m,1}$}} $=$ (0,0,0) and \textbf{\textit{k$_{m,2}$}} $=$ (0,0.5,0) with respect to the monoclinic $P$2 cell, described in the standard setting of the monoclinic space group. Models using (0.5,0,0) and (0,0,0.5) clearly show that not all magnetic peaks are fit (see Figure S1 in Supplemental Material\cite{Supplemental_Material} and references \cite{Tragheim134} therein). Therefore, \textbf{\textit{k$_2$}} is found to lie along the direction of the long JT bond in the Mn$^{3+}$O$_6$ octahedra, and hence along the same direction where the $G$-type orbital order, tetragonal strain with respect to the $Im\bar{3}$ $AA'_3B_4$O$_{12}$ aristotype and ferroelectric polarization distortions act. The magnetic peak at $d$ $=$ 6.7\,\AA{} is not fit by any of these models and is considered to be a magnetic impurity due to $\alpha$-Mn$_2$O$_3$.

\begin{figure}[t]
\includegraphics{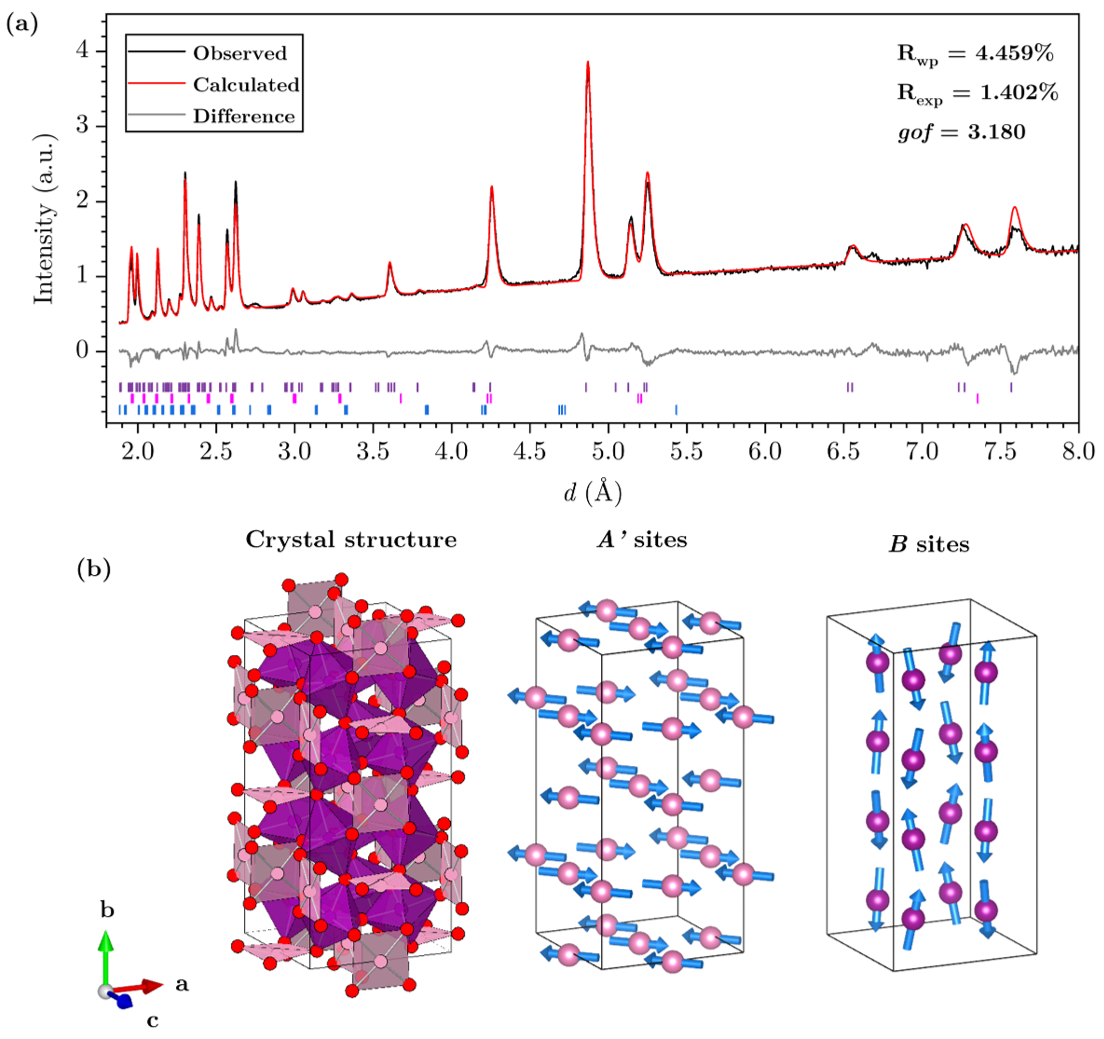}
\caption{\label{Rietveld-model} Rietveld refinement of the resulting $P2_{1}^{'}$ magnetic structure model of Hg$_{0.7}$Na$_{0.3}$Mn$_3$Mn$_4$O$_{12}$ against PND data at 1.5 K. Purple ticks are crystal structural and magnetic peak positions of the fitted model. Pink and blue ticks are crystal structural peak positions due to small amounts of the higher temperature $R\bar{3}$ phase ($\sim$3$\%$) and Mn$_2$O$_3$ ($\sim$3$\%$), respectively. Fitting statistics and goodness of fit ($gof$) are given. (b) The resulting magnetic structure model from refinements in (a). Schematics of the crystal structure, magnetic moments of the $A'$ sites and magnetic moments of the $B$ sites are given separately for clarity. Red spheres are oxygen, $A$ sites are omitted for clarity, and magnetic moments are given as blue arrows.} 
\end{figure}

By systematically testing models derived from the $P$2 nuclear structure and adding magnetic distortions with the \textbf{\textit{k$_{m,1}$}} $=$ (0,0,0) and \textbf{\textit{k$_{m,2}$}} $=$ (0,0.5,0) propagation vectors\cite{Isotropy}, it is found that only a model with a $P2_{1}^{'}$ magnetic space group, defined in a unit cell obtained through the transformation matrix \{(1,0,0),(0,2,0),(0,0,1)\} and origin (0,0,0), is able to give a good fit of the data with physical values of the ordered moments on the Mn sites. Figure \ref{Rietveld-model} gives the Rietveld refinement of the $P2_{1}^{'}$ magnetic structure model against PND data at 1.5\,K. The resulting magnetic moment values for each site and direction are given in Table \ref{moments}.

\begin{table}[b]
\caption{\label{moments} Magnetic moment direction and total magnitudes of the Mn $A'$ and $B$ sites in Hg$_{1-x}$Na$_x$Mn$_3$Mn$_4$O$_{12}$ at 1.5\,K from refinements given in Figure \ref{Rietveld-model}. Estimated standard deviations (1$\sigma$) are given in parentheses.}
\begin{ruledtabular}
\begin{tabular}{ccc}
\textrm{Site}&
\textrm{Moment direction}&
\textrm{Magnetic moment ($\mu_\text{B}$)}\\
\colrule
Mn$^{A'}$ & \textbf{\textit{a}} $=$ \textbf{\textit{c}} & 1.857(11) \\
Mn$^{A'}$ & Magnitude & 2.626(16) \\
Mn$^{B}$ & \textbf{\textit{a}} $=$ \textbf{\textit{c}} & 0.40(2) \\
Mn$^{B}$ & \textbf{\textit{b}} & 2.649(12) \\
Mn$^{B}$ & Magnitude & 2.71(3) \\

\end{tabular}
\end{ruledtabular}
\end{table}

For this magnetic structure model, magnetic peaks due to $A'$ site ordering are fit well by refining a single magnetic moment where magnitudes along \textbf{\textit{a}} and \textbf{\textit{c}} are fixed to be equal to each other, as given in Table \ref{moments}. Models that allow for different moment magnitudes along \textbf{\textit{a}} and \textbf{\textit{c}} produce values that are within error of each other ($i.e.,$ within 3$\sigma$) and do not significantly improve the quality of the fit. The moment direction configuration was determined from the symmetry analysis and Rietveld refinements of different models as mentioned above, resulting in a $G$-type-like AFM alignment of moment directions lying within the \textbf{\textit{a}}-\textbf{\textit{c}} plane. By comparison, quadruple manganite perovskites with trivalent $A$-site cations ($A$ $=$ La, Ce, Nd, Sm, Eu, Y) exhibit magnetic structures where $A'$ site moments align in the \textbf{\textit{a}}-\textbf{\textit{c}} plane in a similar manner, but the magnetic ordering is ferrimagnetic instead of $G$-type-like compared to the model reported here\cite{johnson2018magnetic, johnson2019displacive}. Hence, some similarities may be observed between the $A'$ site ordering between these two cases. A mode decomposition of the final model with respect to the aristotype $Im\bar{3}$ structure indicates that the $A'$ orders only using $mH_4^+$ modes corresponding to the \textbf{\textit{k$_1$}} $=$ (1,1,1) propagation vector with order parameter direction (OPD) $(\delta_1,\delta_2,0)$.

The magnetic peaks due to $B$ site ordering are fit well using only two different components, given in Table \ref{moments}: one describing moment magnitudes along the \textbf{\textit{b}} axis, and a second describing a canting of moments of equal magnitude along \textbf{\textit{a}} and \textbf{\textit{c}}. Here we fixed the moment on the symmetry inequivalent $B$ sites to be constant, however we note that other options, for example an nonphysical like zero moment on one site and double on the other, will fit against the data in the same way. This is to say that we cannot determine the OPD of the magnetic distortion from PND data. The moment alignment along \textbf{\textit{b}} produces an `up--up--down--down' configuration which is responsible for the doubling of the magnetic unit cell compared to the $P$2 crystal structure unit cell. As well as this `up--up--down--down' configuration, the second canting moment is required to fit the data adequately since models that refine only a single or two different moments along \textbf{\textit{b}} do not fit the magnetic peak appearing at $d$ $\approx$ 6.55\,\AA{} (see Figure S2), conclusively indicating that this canting occurs in Hg$_{0.7}$Na$_{0.3}$Mn$_3$Mn$_4$O$_{12}$. This canting is coincident with the direction in which the JT long bond of Mn$^{3+}$O$_6$ octahedra tilt about (see Figure \ref{Rietveld-model}), giving a possible physical origin of the canting observed. The mode decomposition indicates for the $B$ site that the ordering transforms as the irrep $m\Delta_{1}$ with an OPD $(\mu_1,\mu_2;0,0;0,0)$. Similar to the $E$-type AFM order, this ground state magnetic moment configuration on the $B$ site is observed in a number of different perovskite systems including $R^{3+}_2$MnCoO$_6$ ($R$ $=$ Yb, Lu)\cite{blasco2015evidence, blasco2017magnetic}, HoMnO$_3$\cite{munoz2001complex} and various $R^{3+}$NiO$_3$ ($R$ $\neq$ La)\cite{garcia1994neutron, rodriguez1998neutron, fernandez2001magnetic}. In these systems it is generally discussed that the $E$-type AFM order breaks inversion symmetry and results in potential ferroelectric polarization\cite{sergienko2006ferroelectricity}. For the $R$MnO$_3$ manganites, the crossover from $A$-type to $E$-type magnetic order as the size of the $R^{3+}$ cation decreases from La$^{3+}$ to Lu$^{3+}$ has been attributed to an increased degree of octahedral rotations, leading to a frustration between 180° AFM and 90° FM superexchange.\cite{solovyev2009long} The octahedral rotation angle in Hg$_{0.7}$Na$_{0.3}$Mn$_3$Mn$_4$O$_{12}$ is on average 138°, compared to 144° in HoMnO$_3$, suggesting that this may be a factor here too in stabilizing the `up--up--down--down' magnetic structure. Indeed, a spin-induced polarization may be observed in Hg$_{0.7}$Na$_{0.3}$Mn$_3$Mn$_4$O$_{12}$, similar to that observed in our previous measurements on HgMn$_3$Mn$_4$O$_{12}$ which demonstrate switchable pyrocurrent and polarization signatures at a temperature coincident with the onset of magnetic order\cite{chen2018improper}. We will show later that the coupling of these magnetic modes to the improper polarization could explain its stabilization.

\begin{figure}
\includegraphics{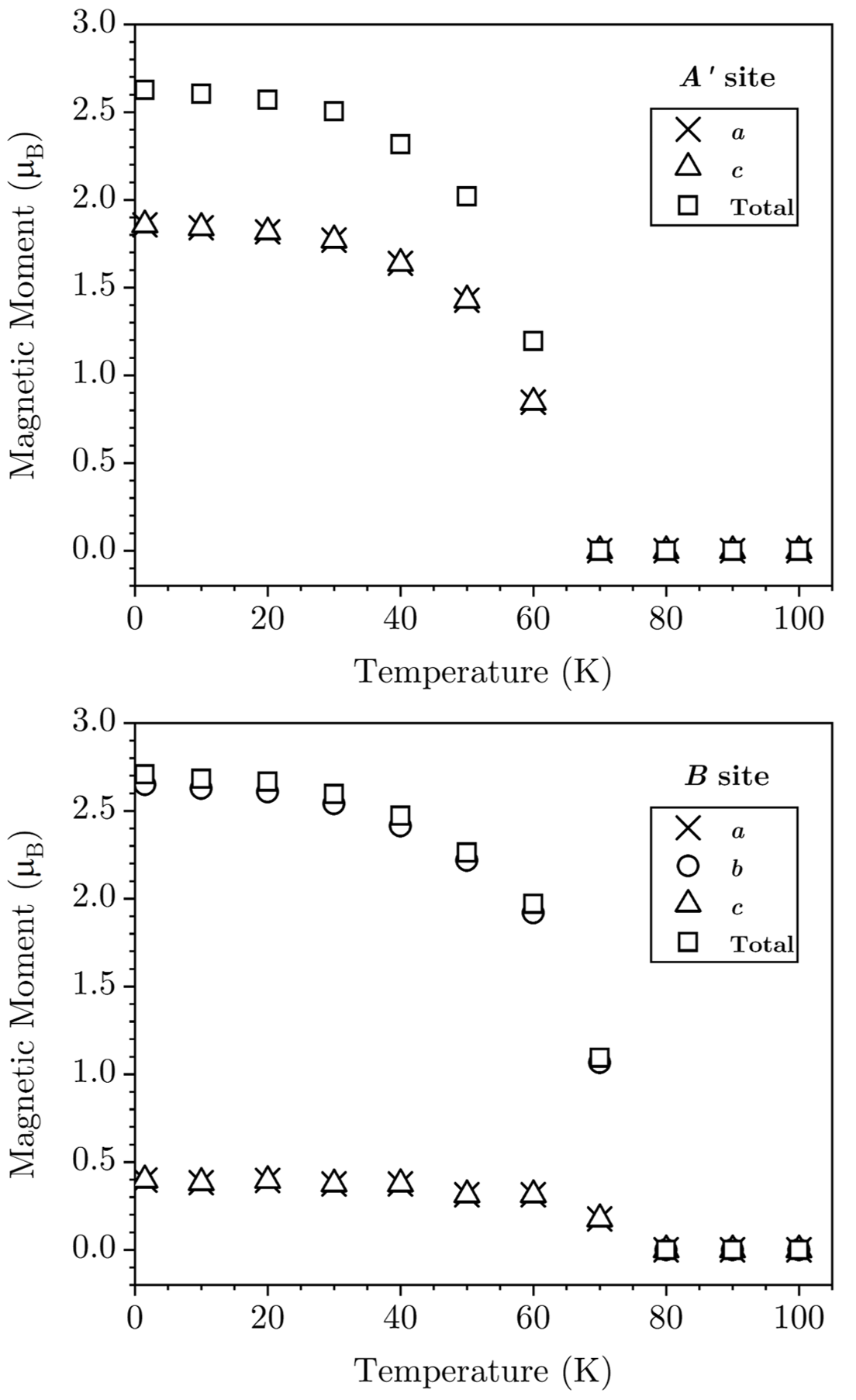}
\caption{\label{VT-magnetic-modes} Temperature dependence of the magnetic moments for $A'$ (upper) and $B$ (lower) sites for the $P2_{1}^{'}$ magnetic structure model. Estimated standard deviations (1$\sigma$) are smaller than the size for each data point, and so they are not given here.}  
\end{figure}

Figure \ref{VT-magnetic-modes} illustrates the temperature evolution (see Figure S3 for raw data) of each magnetic moment component, and total moment, for both $A'$ and $B$ sites. The temperature onset for $B$ site ordering ($\lesssim$ 80\,K) is shown to occur at a higher temperature compared to $A'$ site ordering ($\lesssim$ 70\,K), consistent with magnetic susceptibility data in Figure \ref{magnetic-susceptibility}. The fact that the two sublattices order at different temperatures with two different propagation vectors and perpendicular direction of the spins strongly indicates that the inter-sublattice interactions are weak, effectively making the two independent from each other.  

\subsection{Symmetry consideration on the $B$ site magnetic ordering}\label{Symmetry}

\begin{figure}
\includegraphics{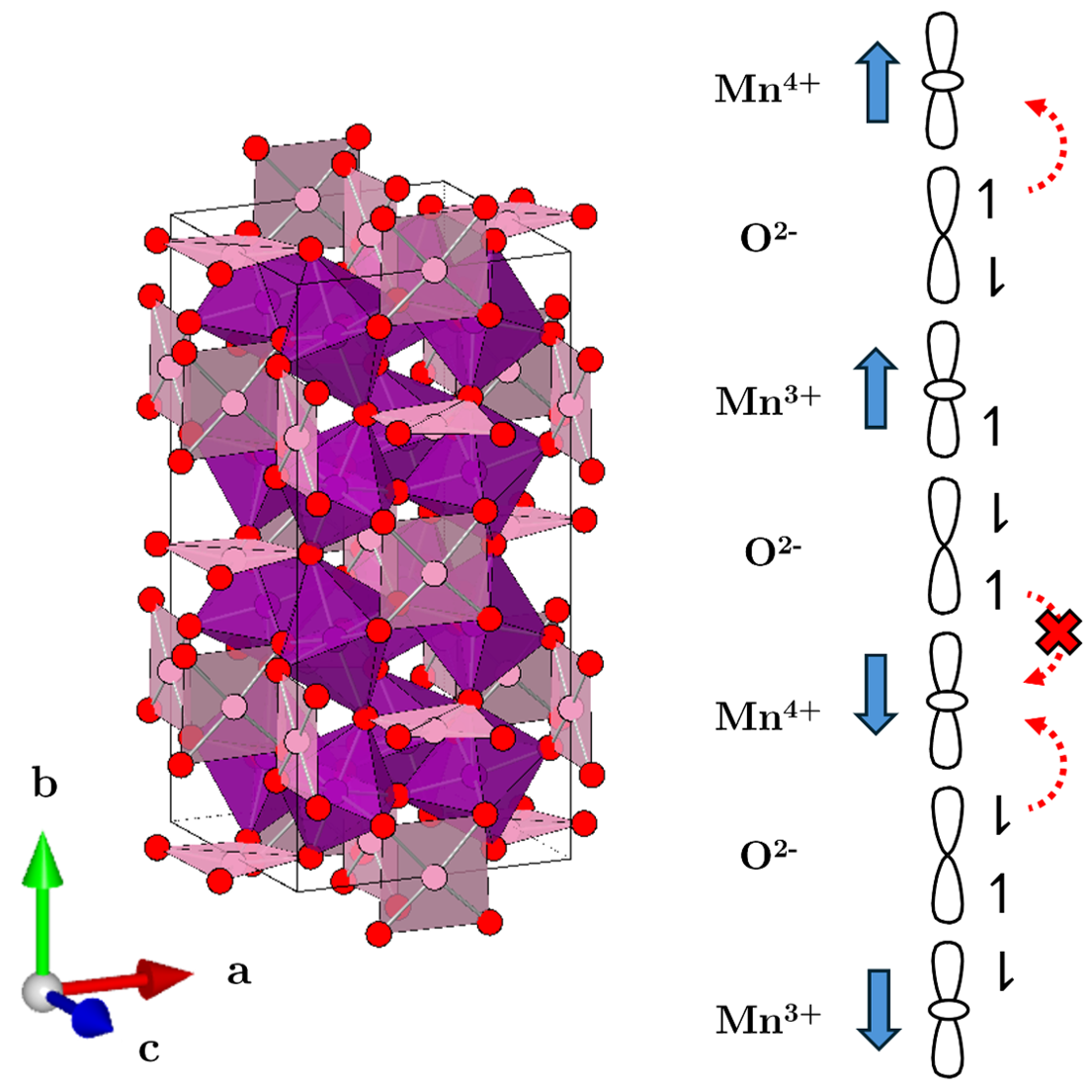}
\caption{\label{mechanism} Proposed exchange mechanism schematic of the $B$ sites for the $P2_{1}^{'}$ magnetic structure of Hg$_{0.7}$Na$_{0.3}$Mn$_3$Mn$_4$O$_{12}$. $d_{z^2}$ orbitals for Mn$^{3+}$ and Mn$^{4+}$, and the 2$p$ orbital for O$^{2-}$ are given. Dark blue arrows are the magnetic moment directions for an arbitrary `chain' of $B$ cations along the \textbf{\textit{b}} unit cell direction consistent with Figure \ref{Rietveld-model}. Relative spin orientations of each electron for each cation/anion are given for reference, and without canting for clarity.}  
\end{figure}

While a $G$-type magnetic structure on the $B$ site can be understood in terms of a strong AFM near neighbor interaction, the observed `up--up--down--down' configuration is less straightforward to understand given the $G$-type charge and orbital ordering present. Indeed, by examining the magnetic model and exchange mechanism shown in Figure \ref{mechanism} one can appreciate that the double exchange along the \textbf{\textit{b}} axis should result in a FM alignment of Mn$^{3+}$ and Mn$^{4+}$ moments, mediated by the $G$-type orbital order and tetragonal strain acting along this direction. Considering this interaction, it appears that every other bond in the structure will be frustrated (see Figure \ref{mechanism}). One could expect that this frustration could be alleviated through a Peierls distortion where the FM coupled Mn $B$-site cations displace towards each other, resulting in a Zener polaron-like behavior. This kind of Peierls distortion would then be expected to produce a slight perturbation of the site-centered rocksalt charge order on the $B$ sites and promote some bond-centered charge order character, similar to previously proposed models of Zener polaron order in some Pr$_{1-x}$Ca$_{x}$MnO$_3$ compositions for $x$ $<$ 0.5\cite{daoud2002zener, wu2007experimental}.

\begin{figure}
\includegraphics{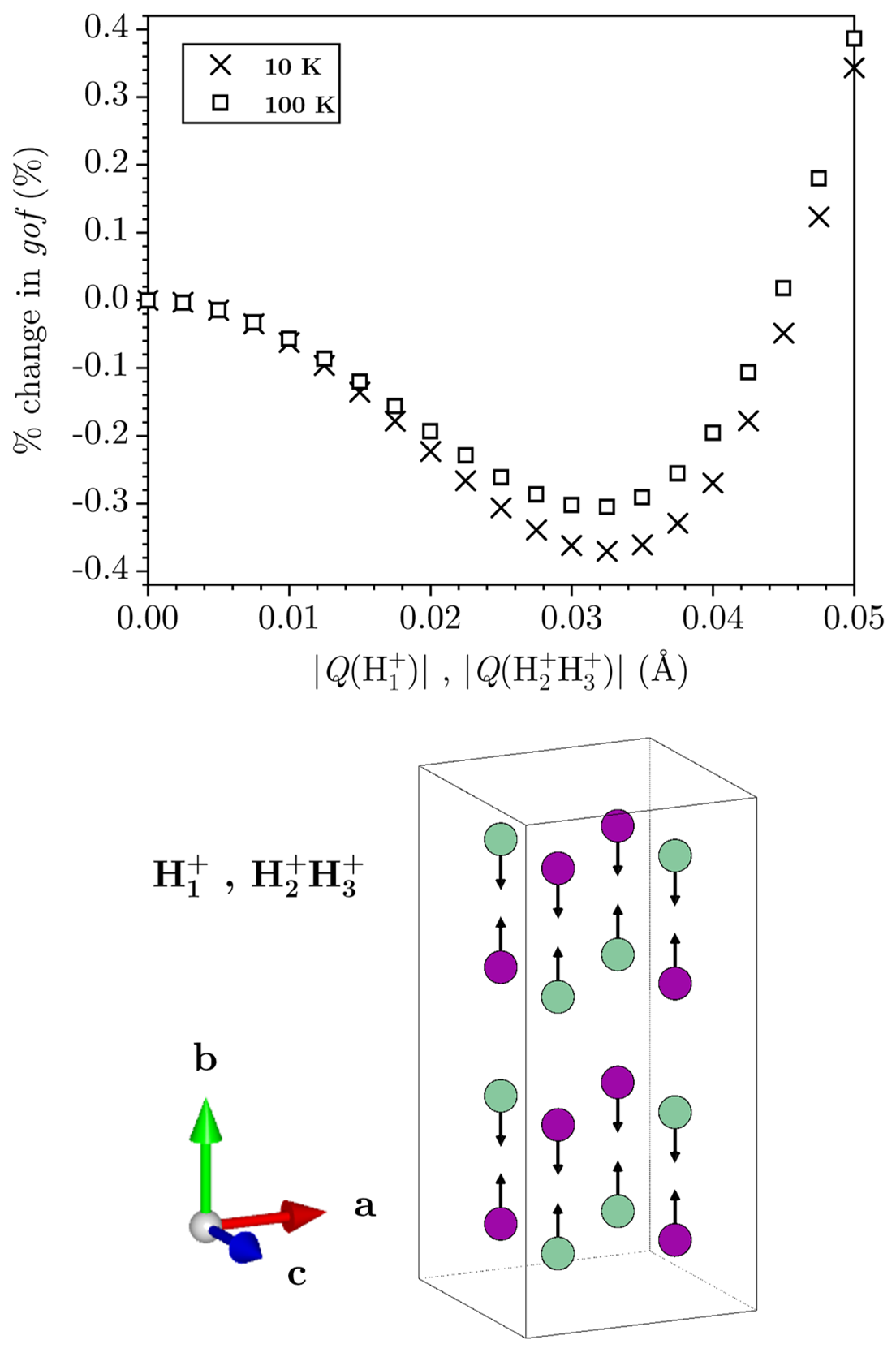}
\caption{\label{H-point-modes-refinements} Percentage changes in the goodness of fit ($gof$) for refinements of models against previously collected synchrotron PXRD data, collected at 10\,K and 100\,K, as a function of tuning Zener polaron-like Peierls structural distortion modes H$_1^+$ and H$_2^+$H$_3^+$. The bottom figure illustrates schematically the action of both H$_1^+$ and H$_2^+$H$_3^+$ on the $B$ sites. Here, purple and green circles illustrate the rocksalt charge order of Mn$^{3+}$ and Mn$^{4+}$ occurring in the crystal structure. All other sites are omitted for clarity.} 
\end{figure}

In our previous structural characterization of Hg$_{1-x}$Na$_x$Mn$_3$Mn$_4$O$_{12}$ our refinements were performed allowing degrees of freedom related to the site-centered $B$ site rock-salt charge order of $P$2 $G$-type charge and orbital ordered states to refine freely down to 10\,K\cite{Tragheim134}. The structural distortion that results in this $G$-type charge and orbital order transforms as the irrep H$_2^-$H$_3^-$ with an OPD $(\rho_1,\rho_2)$, with respect to the $Im\bar{3}$ $AA'_3B_4$O$_{12}$ quadruple perovskite aristotype. This irrep describes oxygen-based displacements, and the corresponding modes were refined in our previous refinements. On the other hand, the Zener polaron-like Peierls distortions of Mn $B$ sites that occur along the \textbf{\textit{b}} axis, and collectively transform as the irreps H$_1^+$ with OPD $(\chi)$ and H$_2^+$H$_3^+$ with OPD $(\phi_1,\phi_2)$, were fixed as zero in our original refinements. Here, by revisiting our synchrotron high-resolution PXRD data\cite{Tragheim134} we perform a series of refinements of models against these data at temperatures $T$ $=$ 10\,K and 100\,K, capturing temperatures both above and below the onset of magnetic order. In these refinements, mode amplitudes of H$_1^+$ and H$_2^+$H$_3^+$ are systematically increased, but fixed, by small and equal increments while the mode amplitudes of all other crystal structure distortion modes are allowed to freely refine. This refinement procedure then allows the determination of whether the quality of fits are improved as a function of H$_1^+$ and H$_2^+$H$_3^+$. In other words, whether forcing additional Peierls-like distortion modes to our structural models provides a mechanism through which the magnetic structure forms in Hg$_{0.7}$Na$_{0.3}$Mn$_3$Mn$_4$O$_{12}$. The quality of fits of these refinements are measured by percentage changes in the goodness of fit ($gof$) in systematically tuning the magnitude of H$_1^+$ and H$_2^+$H$_3^+$ modes, $|$$Q$(H$_1^+$)$|$ and $|$$Q$(H$_2^+$H$_3^+$)$|$, from zero to non-zero values. The results from this analysis are given in Figure \ref{H-point-modes-refinements}. 

It is observed from these refinements that at temperatures both above and below magnetic ordering there exists a consistent but small ($\sim$ 0.4$\%$) minima for the percentage difference in $gof$ for non-zero $|$$Q$(H$_1^+$)$|$ and $|$$Q$(H$_2^+$H$_3^+$)$|$. This minima occurs for these Zener polaron-like distortion mode values of $\sim$ 0.03\,\AA{}, however this is $\sim$ 3--6$\times$ smaller than the refined $G$-type charge/orbital order distortion mode values (Figure S4 and S5). This means that while these Zener polaron-like distortions are allowed by symmetry in the $P$2 crystal structure, their small amplitudes suggest that they arise due to an improper coupling to other distortion modes active in $G$-type charge and orbital ordered Hg$_{1-x}$Na$_x$Mn$_3$Mn$_4$O$_{12}$. Furthermore, since these minima are observed both above and below magnetic ordering, this suggests that these Zener polaron-like distortions are not induced by the magnetic order parameter. Nevertheless it is possible to derive a free energy invariant between the H$_2^+$H$_3^+$ and m$\Delta_1$ modes, $A(\phi_1\mu_1^2-\phi_1\mu_2^2)+B(\phi_2\mu_1^2-\phi_2\mu_2^2) $ where $A$ and $B$ are phenomenological constants, which could explain the stabilization of the `up--up--down--down` magnetic structure. As highlighted before, the amplitudes of the H$_2^+$H$_3^+$ modes are very small, meaning that the energy gain from this coupling is also very small.

Due to this small energy gain offered from the coupling to the Zener polaron-like distortion we investigated other possible coupling terms between the magnetic order parameter and the displacive distortions present in the paramagnetic state. To summarize, the nuclear $P$2 structure is obtained from the parent $Im\bar{3}$ aristotype by the action of two main primary distortions transforming as the irreps H$_{4}^{+}$ with an OPD $(0,0,\sigma)$ and H$_{2}^{-}$H$_{3}^{-}$ with an OPD $(\rho_1,\rho_2)$. These two primary order parameters induce the electrical polarization, transforming as $\Gamma_{4}^{-}$ with an OPD $(0,\tau,0)$, through the trilinear coupling $A(\sigma\tau\rho_1 ) + B(\sigma\tau\rho_2)$ where $A$ and $B$ are phenomenological constants. While a direct coupling between the magnetic order parameter and any of the H point primary displacive distortions is not allowed by symmetry, it is possible to find a direct coupling between the $m\Delta_1$ magnetic order parameter and the induced electrical polarization. This coupling term is a $5^{th}$ degree invariant of the form $\tau (\mu_{1}^{3}\mu_{2} - \mu_{1}\mu_{2}^{3})$ and acts as a lock-in invariant which locks the \textbf{\textit{k$_2$}} propagation vector to the (0,0.5,0) value since this is not a special point of the Brillouin zone for the $Im\bar{3}$ space group. A similar effect of the electrical polarization in selecting the magnetic ground state has been observed in BiMn$_{7}$O$_{12}$ where the proper Bi polarization selects the `up--up--down--down' magnetic structure through the inverse exchange striction mechanism \cite{Behr2023}. For BiMn$_{7}$O$_{12}$, the proper Bi polarization with its large amplitude and the low degree of the coupling invariant will provide a large stabilization factor, whereas in Hg$_{0.7}$Na$_{0.3}$Mn$_3$Mn$_4$O$_{12}$ the polarization is improper in nature with a higher degree coupling invariant.  

In this section, we showed that two different coupling mechanisms between the magnetic order parameters and two secondary nuclear distortions exist by symmetry in the free energy of the system. Since the nuclear distortion amplitudes involved in the couplings are small, and due to their improper nature, it is unlikely that these can be considered the main driving mechanism for the `up--up--down--down' structure. Nevertheless, we speculate that these terms could tip the energy balance in favor of such a structure. Indeed, if we consider a simple Heisenberg toy model on a 1D chain along the cell \textbf{\textit{b}} axis with a FM near neighbor interaction, as expected from the $G$-type charge and orbital ordering, and a AFM next-near-neighbor we will obtain a frustrated system. The solution of the Hamiltonian will depend on the $J_1$/$J_2$ ratio and could stabilize an incommensurate non-collinear phase. In this situation the free energy terms we described above could push the system to lock the propagation vector to the (0,0.5,0) position and in this way obtain the `up--up--down--down' ground state. More theoretical investigations are needed to either confirm or disprove this speculation, but these are beyond the scope of the current work.

\section{Conclusion}

We present the ground state magnetic structure of the quadruple manganite perovskite Hg$_{0.7}$Na$_{0.3}$Mn$_3$Mn$_4$O$_{12}$, exhibiting the novel $G$-type charge and orbital ordered state in comparison to other $A^{2+}$Mn$_3$Mn$_4$O$_{12}$ ($A$ $=$ Ca, Sr, Cd, Pb) systems. The magnetic structure is found to exhibit two magnetic transition temperatures on decreasing temperature attributed to ordering of Mn on the $B$ and $A'$ sites. The $A'$ site demonstrate a $G$-type-like AFM order constrained within the \textbf{\textit{a}}-\textbf{\textit{c}} plane, and the $B$ sites demonstrate a canted `up--up--down--down' AFM order with primary moment alignments along the \textbf{\textit{b}} axis. We investigate the mechanisms driving the magnetic ordering of the $B$ sites from a symmetry perspective highlighting two possible coupling mechanism with improper nuclear distortions. We speculate that these two terms could push the system from an incommensurate state, resulting from the frustration between near and next near neighbor, to the commensurate `up--up--down--down' structure. 

The differences in the magnetic structure of Hg$_{0.7}$Na$_{0.3}$Mn$_3$Mn$_4$O$_{12}$ compared to magnetic structures reported for other $A^{2+}$Mn$_3$Mn$_4$O$_{12}$ systems further highlights the intriguing and distinct properties of Hg-quadruple manganite perovskites in comparison, attributable to the unique chemistry of Hg$^{2+}$ residing on the $A$ site. The results presented here will be of significant interest to those who wish to study the complex interplay between exotic charge, orbital, electronic and magnetic orders in strongly correlated systems.      

\begin{acknowledgments}
B. R. M. T thanks the University of Warwick and the EPSRC for studentship and funding (EP/R513374/1) and M. S. S acknowledges the Royal Society for a fellowship (UF160265 \& URF\textbackslash R\textbackslash231012). W.-T. C thanks the supports from National Science and Technology Council in Taiwan (Grants Nos. 111–2112-M-002–044-MY3, 113–2124-M-002–006, and 114-2112-M-002-030), the Featured Research Center Program within the framework of the Higher Education Sprout Project by the Ministry of Education in Taiwan (Grants No. 113L9008), and Academia Sinica (Project No. AS-iMATE-115-14). Experiments at the ISIS Neutron and Muon Source were supported by beamtime allocation RB2400051 from the Science and Technology Facilities Council. The experimental data that support these findings are available\cite{WISH_data}. 
\end{acknowledgments}

\bibliography{references}

\end{document}